# Influence of the Working Gas Properties on the Anode Wire Modulation Effect of MWPC

WANG Xiao-hu

(Fundamental Science on Nuclear Waste and Environmental Security Laboratory, Southwest University of Science and Technology，Mianyang，621010)

**Abstract**   For MWPC used for X-ray position detection, simulation study of the anode wire modulation effect of the detector was carried out with Garfield program. Different gas mixtures were used as the working gas in the simulation, so as to obtain the influence of the X-ray cross section and electron diffusion coefficient of the working gases on the anode wire modulation effect of MWPC which has a anode wire spacing of 2mm. Results show that, though working gas with higher X-ray cross section implies a larger average drift distance of ionized electrons, using of such gas mixtures is of little use to improve the anode wire modulation effect of MWPC. And the transverse electron diffusion coefficient is the determining factor that affects the extent of the anode wire modulation effect of the detector.

## 1 Introduction

Position sensitive detectors based on Multi-wire Proportional Chambers (MWPCs) are widely used in neutron & X-ray detection. MWPC own many advantages for particle detection, such as stability, cheap, large area can be made, etc. However, because the anode plane of MWPC is made of a set of parallel wires, the anode wire modulation effect is a main factor that limits the spatial resolution in the direction across the anode wires of the detector.

There are many factors that affect the magnitude of anode wire modulation effect of MWPC, including anode wire spacing, drift distance of the ionized electrons, and electric field in the drift region, etc. As mentioned in Ref[1,2], by applying a cathode wire plane with wires parallel and in registration with the anode wires, the anode wire modulation effect and be reduced significantly. Results in Ref[2, 3] also show that, the deeper the drift region is, the smaller anode wire modulation effect the detector performs. The simplest way to obtain larger drift region depth is to increase the thickness of the working gas volume, but a thick working gas volume implies much more parallax error for planar type gas detectors. For X-ray detection, since the attenuation of the X-ray beam in matter obeys the $I=I_0e^{-ud}$ rule, most X-ray will be absorbed near the entrance window if working gas with high X-ray cross section was used, and this means most ionized electron will experience a longer drift distances than these in working gas with low X-ray cross sections. Therefore, applying working gas with high X-ray cross section might be a way to reduce the anode wire modulation effect of MWPC. Meanwhile, since the diffusion of the electrons can affect the sharing of ionized electrons on the anode wires, working gas with different diffusion coefficient corresponds to different anode wire modulation effect of MWPC. Till now, few works have focused on the influence of these gas properties on the anode wire modulation effect of MWPC.

In this paper, the anode wire modulation effect of a MWPC with anode wire spacing of 2mm was studied. The influence of the electron diffusion coefficient and X-ray cross section on the anode wire modulation effect of the detector was investigated.

## 2 Method

The anode wire modulation effect of MWPC arises from the fact that all avalanches caused by an incident particle can only occurs around an anode wire. Therefore, no matter what the actual position the particle owns, its measured position along the axis across the anode wires is always the position of the corresponding anode wire. When applying a uniform irradiation, the measured positions of the incident particles of detector will not be a uniform distribution, but with most counts focus on the locations of the anode wires. So the uniform irradiation response of the detector can be used as the reference of the anode wire modulation effect of MWPC.

In this paper, we use Garfield[4] software to simulate the uniform X-ray irradiation response of a MWPC. Though it is known that MWPC with small anode wire spacing suffers low degree of anode wire modulation effect, the small anode wire spacing implies much more difficulties in detector manufacture. In this work, a MWPC with common used anode wire spacing of 2mm was studied, since if the anode wire modulation can be eliminated at this anode wire spacing it obviously can be eliminated at anode wire spacing smaller than 2mm. The structure of the MWPC is shown in Fig.1. It mainly consists of a drift region and an amplification region. The depth of the drift region is 20mm. The anode plane and the cathode wire plane both have a wire pitch of 2mm, and the distances between the anode plane to the nearest two electrode planes are 2mm too. The diameter of the anode wire and cathode wire used in the simulation is 20μm and 50μm respectively.

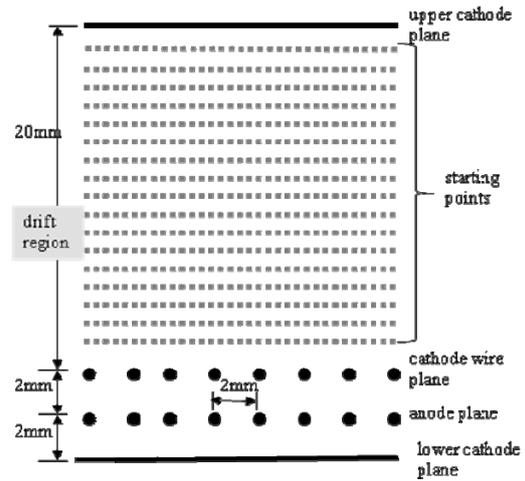

Fig.1 Structure of the MWPC

In the simulation, a series of starting point were uniformly selected in the drift region of the MWPC. At each starting point, 200 electrons were let drift towards the anode plane, so as to simulate an X-ray with energy of several keV being absorbed in the starting point. Because different gas mixture has different X-ray cross section, the number of the simulated X-rays was not the same for all starting points, but was differ from each other according to the distance between the starting points to the entrance window (upper cathode plane). Gas mixtures including Ar/CH4(90/10), Ar/CO2(70/30), Xe/CO2(70/30), Ar/CO2(90/10) and Xe/CO2(90/10) was used in the simulation, since they have different electron diffusion coefficients and X-ray cross sections. Fig.2 is the calculated electron transverse diffusion coefficient obtained by Garfield simulation. As shown in the figure, for electric field ranged from 200V/cm to 5000V/cm, Ar/CH4(90/10) has the largest electron diffusion coefficient, then is Ar/CO2(90/10) and Xe/CO2(90/10), and lowest one is Ar/CO2(70/30) and Xe/CO2(70/30) .

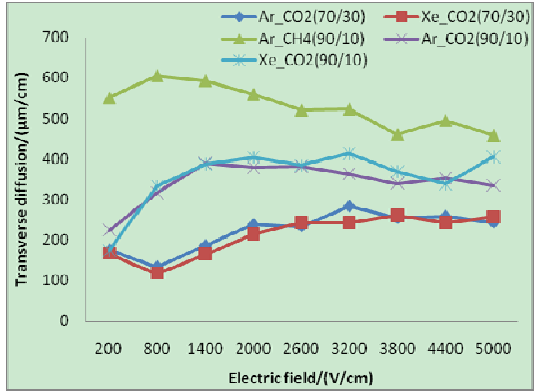

Fig.2 Transverse diffusion of the gas mixtures

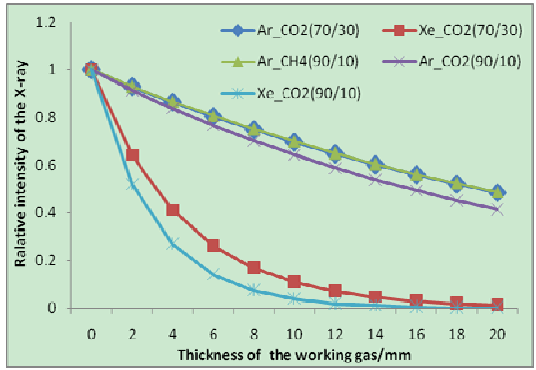

Fig.3 X-ray attenuation in different gas mixtures

Using X-ray cross section database XCOM[5], the attenuation of the 5.9keV X-ray in different working gas can be calculated, and the result is shown in Fig.3. According to Fig.3, xenon based gas mixture has a much larger attenuation coefficient than argon based gas mixture. For Xe/CO2(90/10) gas mixture, more than 96% of the incident X-ray will be absorbed in the upper half of the drift region shown in Fig.1, while the value is only 70% for Ar/CH4(90/10). When applying a uniform X-ray irradiation, the mean electron drift distance in the drift region is 17mm for Xe/CO2(90/10), comparing to that of 11mm for Ar/CH4(90/10).

In the anode modulation effect simulation, primary electrons was let drift from the starting points in the drift region. But the number of the events (an event represents a simulated X-ray, which coresponds to 200 primary electrons be generated at the starting point) at each starting point is chosen according to the curve shown in Fig.3, that is, the closer the starting point to the entrance window the more events it has. The Drift_MC_Electron subrutine in the Garfield program was used to track the ending location of each electron, and the measured position of a simulated X-ray was obtained by calculating the center of garvity of the electron position distribution on the anode plane. By recording the measured position of simulated X-rays form all starting points, the uniform irradiation response of the detetor can be obtained.

## 3 Simulation Results

### 3.1 Optimization of the drift field

The electric field in the drift region is mainly determined by the upper cathode voltage of the MWPC. As mentioned in Ref[2,3], the anode wire modulation effect is strongly affected by the electric field in the drift region. For a given structure and working gas of MWPC, there is an optimized drift electric field that detector obtains the best linearity when operated under this electric field. Fig.4 is the calculated uniform irradiation response (UIR) of the MWPC under different upper cathode voltages. The working gas use in Fig.4 is Ar/CH4(90/10), as can be seen, the measured position distribution is almost uniform when the voltage of the upper cathode is set to -1000V, but the response linearity get worse either at a lower or higher cathode voltages. Therefore, in the simulation below, in order to get the best uniform irradiation response (UIR) of the detector, the drift field (cathode voltage) was adjusted and optimized at first.

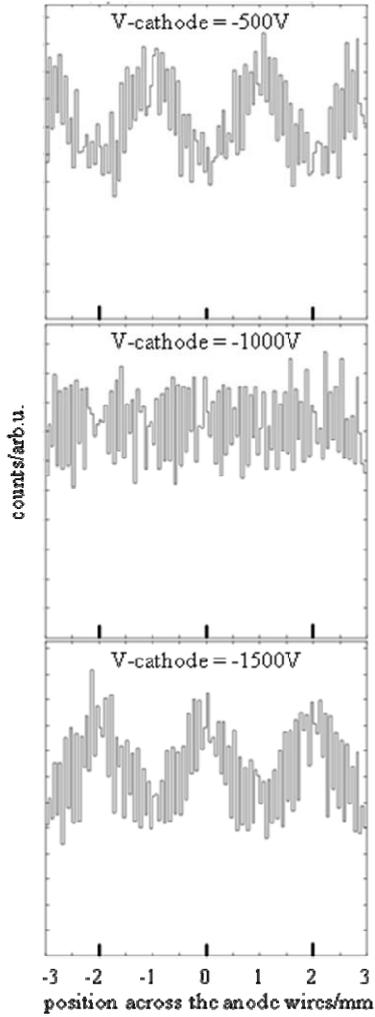

Fig.4. UIR of the detector under different cathode voltages, the anode wires are located at 0, ±2mm respectively, as marked in the Figure.

**3.2 Best UIR for different working gases**

In order to investigate the anode wire modulation effect under different drift region depth, best uniform irradiation response (UIR) of the MWPC with drift region of 20mm and 12mm were calculated. For each working gas, the best UIR of the detector was obtained by adjusting the upper cathode voltages of the MWPC. By our calculation, the optimized drift field for different gas mixture is ranged of 400V/cm to 1100V/cm.

For drift region depth of 20mm, the calculated best UIR of the detector using different working gas was shown in Fig.5. As can be seen, the best UIR of the detector using $Ar/CH_4(90/10)$、$Ar/CO_2(90/10)$ and $Xe/CO_2(90/10)$ is almost a uniform distribution, which means there is almost no anode wire modulation effect in this condition. However, for working gas of $Ar/CO_2(70/30)$ and $Xe/CO_2(70/30)$, the best UIR of the detector still shows a modulation. Though $Xe/CO_2(70/30)$ has larger X-ray cross section, the degree of anode wire modulation effect is higher in $Xe/CO_2(70/30)$ than in $Ar/CO_2(70/30)$, but the period of the modulation is reduced to half of the anode wire pitch.

Fig.6 is the calculated best UIR of the detector using different working gas with drift region depth of 12mm. From Fig.6, we can see that the best UIR of the detector is quite similar to the situation for drift region depth of 20mm. That is, there is almost no modulation for detector using $Ar/CH_4(90/10)$、$Ar/CO_2(90/10)$ and $Xe/CO_2(90/10)$ as working gas. When using $Ar/CO_2(70/30)$ and $Xe/CO_2(70/30)$ as working gas, the anode wire modulation effect still exist, but with the period of the modulation being reduced to half of the anode wire pitch.

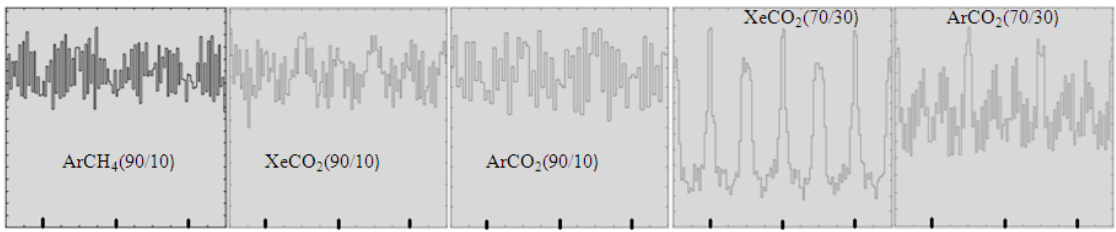

Fig.5 Best UIR spectrum of the detector using different working gases with drift region depth of 20mm, the markers represent for the positions of the anode wires.

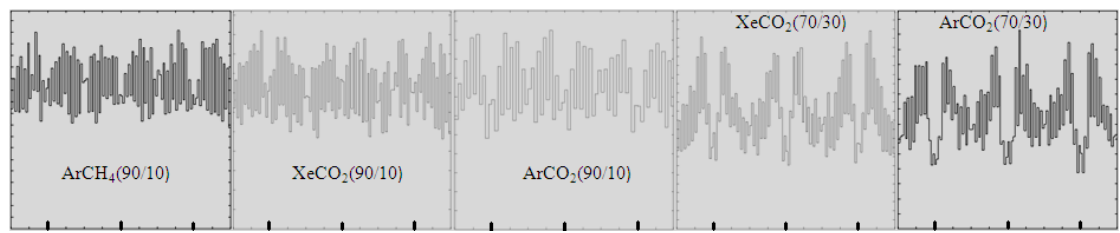

Fig.6 Best UIR spectrum of the detector using different working gases with drift region depth of 12mm, the markers represent for the positions of the anode wires.

## 4 Discussions and Conclusions

Among the gas mixtures, Ar/CO2(70/30) and Xe/CO2(70/30) have the lowest electron diffusion coefficient. Results shows that, even applying optimized drift field, detector using these two gas mixtures cannot eliminate the effect of anode wire modulation. Xe/CO2(90/10) owns the largest X-ray cross section, but it also has a larger diffusion coefficient than Xe/CO2(70/30) and Xe/CO2(70/30). Since detector using Ar/CO2(90/10) or Xe/CO2(90/10) can all get a uniform distribution of UIR shown in Fig.5 and Fig.6, we can deduce that it is the higher electron diffusion coefficient rather than the larger X-ray cross section of Xe/CO2(90/10) that diminish the anode wire modulation effect of the detector.

For cell structure shown in Fig.1, which has an anode wire plane and a cathode wire plane both with a wire pitch of 2mm, by comparing the best X-ray UIR of the three group of gas mixtures (ArCH4(90/10), Ar/CO2(70/30) and Xe/CO2(70/30), Ar/CO2(90/10) and Xe/CO2(90/10) ), following conclusion can be made.

(1) Working gas with high X-ray cross section is almost of little use to eliminate the effect of anode wire modulation. Though Xenon based gas mixtures own larger X-ray cross section than Argon based gas mixture, the anode wire modulation effect of the detector is almost the same for these two types of gas mixture if their transverse electron diffusion coefficient is identical.

(2) Detector with drift region depth of 20mm and 12mm performs quite similar best uniform irradiation response, which means that the anode wire modulation effect cannot be effectively reduced by increasing the drift region depth.

(3) Transverse diffusion coefficient of the working gas is the determinant factor that influences the extent of the anode wire modulation of the detector. Detector with working gas of Ar/CH4(90/10)、Ar/CO2(90/10) and Xe/CO2(90/10) can get a nearly uniform position response when applying a simulative uniform X-ray irradiation. But for detector using Ar/CO2(70/30) and Xe/CO2(70/30) the anode wire modulation still cannot be eliminated. Considering the difference of the transverse electron diffusion coefficient of the gas mixtures shown in Fig.2, we can deduce that if the average transverse diffusion coefficient can be kept larger than about 300μm/cm at field from 400V/cm to 1100V/cm the anode wire modulation effect can be eliminated.